\documentclass[aps,prl,twocolumn,amsmath,amssymb,floatfix]{revtex4}

\usepackage{revsymb}
\usepackage{graphicx} 
\usepackage{amsmath}
  \usepackage{tikz}

\usepackage{setspace}

\DeclareGraphicsExtensions{.jpg,.png}

\begin{document}

\title{Mass crystals and cage effect in vorticity crystals}

\author{Jean-R\'{e}gis Angilella\footnote{email: Jean-Regis.Angilella@unicaen.fr}}
\address{Normandie Universit\'e, UNICAEN, UNIROUEN, ABTE, ESIX Cherbourg, Caen 14000, France.}

\begin{abstract}
{
We study the motion of tiny heavy inertial particles advected by a two dimensional inviscid fluid flow composed of $N$ identical point vortices regularly placed on a ring, and forming a crystal. In the limit of weak particle inertia, we show asymptotically that, in the reference frame of the crystal, inertial particles have $N$ asymptotically stable equilibrium positions located outside the crystal,   in agreement with numerical observations by Ravichandran {\it et al.} (S{\=a}dhan{\=a} 42, 2017). 
In addition to these "satellite" attracting points, we observe that for $N \ge 3$ the center of the ring, though degenerate, is a stable equilibrium position for inertial particles. This creates a kind of cage effect, where inclusions slowly drift towards the center under the effect of the surrounding vortices.  
This cage effect is observed to persist even at larger Stokes numbers, in contrast with the satellite attracting points that vanish when the Stokes number is above some critical value. 
  }
\end{abstract}


\maketitle

 Vortex crystals are two-dimensional flow structures appearing in inviscid fluids, where point vortices place themselves naturally at the tips of polygonal structures that rotate as a solid body \cite{Aref2002review}.  
 They were observed experimentally four decades ago in superfluid Helium \cite{Yarmchuk1979}, then in pure electron columns \citep{Fine1995,Durkin2000PoF,Durkin2000PRL},  floating disks \citep{Grzybowski2000,Grzybowski2002}, and more recently on the poles of Jupiter  \cite{Adriani2018}.  Both experiments and simulations showed that crystals can emerge from random vortex distributions and persist even in the presence of weak viscous effects \cite{Schecter1999,Jin2000,Siegelman2022}. Shortly after the initial disordered state, vortices advect each other in a chaotic manner, and produce vortex layers creating a diffuse background \cite{Schecter1999}. Successive vortex mergers lead to intense vortices  which eventually converge towards a stable configuration in rigid rotation.   
 
 The influence of this self-organization on the transport of inertial particles suspended in the fluid has received little attention so far. However, it can be conjectured that the emergence of such flow structures will certainly affect the distribution of particles, as these objects are known to be very sensitive to the distribution of vorticity. In strong contrast with turbulent flows however,  vortex crystals induce steady flows in the reference frame rotating with the crystal. 
 In the case where two point vortices rotate around each other, their centres forming a rigid segment, it has been shown that heavy inertial particles could be trapped by  attracting  points at rest in the frame of the segment \cite{Angilella2010,Ravichandran2014}.  This property has been observed to persist in the case where more than two vortices interact to form a crystal \cite{Ravichandran2017}. The goal of the present work is to address this question theoretically, in the case of heavy inertial particles transported at low particle Reynolds number in an inviscid potential flow composed of identical vortices regularly placed on a ring. It will be shown that equilibrium points always exist in the frame of the crystal for inertial particles, in the limit of weak inertia, and that these points are hyperbolic (i.e. the real parts of the eigenvalues of the particle flow gradient are non-zero there). Some of these points will be shown to be asymptotically stable, so that the particle cloud at long time takes the form of a mass crystal at rest in the reference frame of the vortex crystal. These trapping points being located away from the center of vorticity, they will be referred to as satellite points. In addition, in the present case where no vortex is placed at the center of vorticity, we will show that this point, though degenerate  (it is a center point with pure imaginary eigenvalues), is asymptotically stable and can attract a significant number of particles. Inclusions trapped there will then be surrounded by the vortices and maintained   within a kind of cage.  In the laboratory frame, these particles will appear as almost motionless at long times, while vortices rotate around them. 
 Both the satellite  attracting  points and the cage effect will be studied asymptotically, in the limit of small  Stokes numbers. We will then show some numerical simulations, involving up to seven vortices, which confirm theoretical predictions.  
  
Prior studying the motion of inclusions in vortex crystals, we recall some basic considerations about equilibrium positions of inertial particles \citep{Babiano2000, Haller2008,Cartwright2010}. 
 We assume particles are much heavier than the fluid, do not interact, and have a small particle Reynolds number. 
Under these conditions, the non-dimensional motion equation of an   inertial particle with time-dependent position $\mathbf X(t)$, transported in a steady flow with velocity field $\mathbf{u(X)}$, assuming a linear drag and neglecting the added mass and Boussinesq-Basset forces, is of the form: 
\begin{equation}
\ddot {\mathbf X} = \frac{1}{St} (\mathbf u(\mathbf X) - \dot {\mathbf X} ) + \mathbf F(\mathbf X,\dot {\mathbf X} )
\label{eqmvt}
\end{equation}
where $St$ is the Stokes number, that is the non-dimensional response time of the inclusion.  In the case of spherical particles with mass $m_p$ and radius $r_p$, the Stokes number reads $St = \Omega_0 m_p /(6\pi \mu \, r_p)$, where $\mu$ is the dynamic viscosity of the fluid and $1/\Omega_0$ is a characteristic flow time   which will be defined below.

The term $ \mathbf F$ contains various terms corresponding to body forces,   hydrodynamic forces or fictitious forces (if the reference frame is non-inertial). 
Equilibrium positions, if any, correspond to 
 \begin{equation}
 \mathbf \Phi(\mathbf X,St) \equiv  \mathbf u(\mathbf X)   + St\, \mathbf F(\mathbf X,  {\mathbf 0} ) = {\mathbf 0} .
   \label{equil}
\end{equation}
We assume that the fluid flow has a stagnation point at some position $\mathbf X_{eq}^0$: $\mathbf u (\mathbf X_{eq}^0) = \mathbf 0$. Then, by virtue of the implicit functions theorem, if the gradient of  
$\mathbf  \Phi$ at $(\mathbf X_{eq}^0,0)$ is invertible, that is if the gradient of
$\mathbf u$ at $\mathbf X_{eq}^0$ is invertible,   there exists a continuously differentiable function $\mathbf X_{eq}(St)$, defined for $St$ in the vicinity of 0, and satisfying   $\mathbf \Phi(\mathbf X_{eq}(St),St)=\mathbf 0$,  and $\mathbf X_{eq}(0) = \mathbf X_{eq}^0$. In other words, a single-valued continuous equilibrium branch persists, inertial particles have equilibrium positions that depend continuously on $St$, and which coincides with   $\mathbf X_{eq}^0$ in the limit of small    inertia. This shows that equilibrium positions of inertial particles exist in the vicinity   of any  fluid equilibrium position with non-zero eigenvalues, in the limit of small Stokes numbers. In the following, we will study the position and stability of these points in the case of a vortex crystal. 

\begin{figure*}
\includegraphics[width=0.95\textwidth]{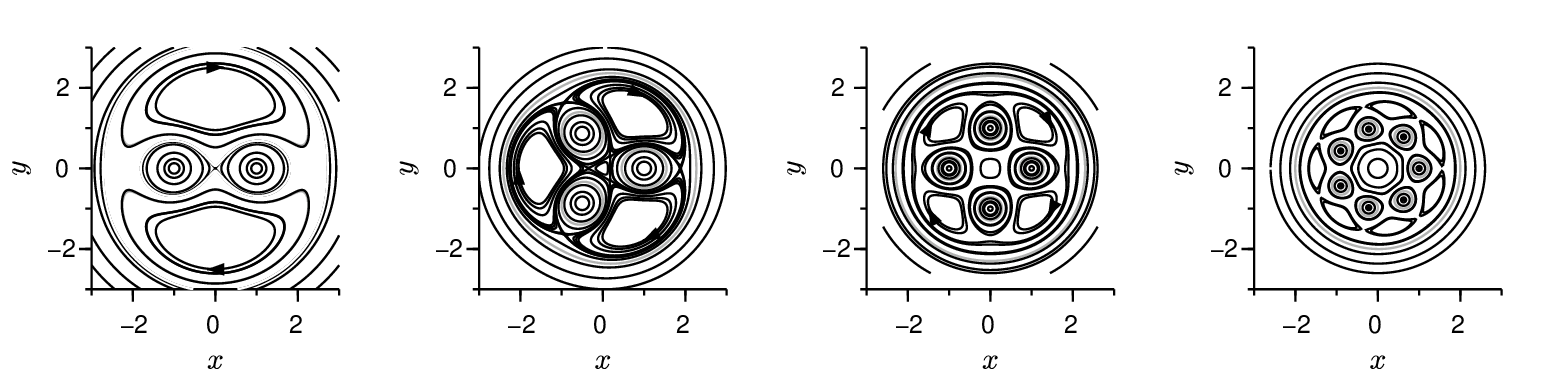}
\caption{Streamlines of the flow in the frame rotating with the crystal, when $N$ = 2, 3, 4 and 7. Lengths are set non-dimensional by means of the radius $a$ of the  circumscribed circle containing vortices. Vortices have a positive sign and are located at $(\cos(n 2\pi/N), \sin(n 2\pi/N))$, $n=0,..,N-1$. In each case, the $N$ recirculation cells are anticyclonic. Note that, when $N > 2$, an additional anticyclonic cell exists around the center-of-vorticity located at $(x,y)=(0,0)$.}
\label{LdC_N2347}       
\end{figure*}

We consider the case where the flow is a vortex crystal composed of $N$ identical point vortices located at $\mathbf x_n = a (\cos(n 2\pi/N), \sin(n 2\pi/N))$, with $n=0,..,N-1$. 
We will limit ourselves to the case $N \le 7$ (Thomson's heptagon),  as a larger number of vortices placed on a ring might lead to unstable flows \cite{Mertz1978}.
These vortices form a rigid body (a regular polygon with radius $a$) that rotates with respect to the laboratory frame, with a constant angular velocity $\Omega_0 = (N-1)\, {\Gamma}/{(4 \pi a^2)}$, 
where    $\Gamma$ is the strength of vortices. 
Without loss of generality we assume $\Gamma > 0$, so that the crystal rotates in the counter-clockwise direction ($\Omega_0 > 0$). 
In the frame rotating with vortices, the flow is steady and its velocity field is given by $\mathbf u = (\partial \psi/\partial y,-\partial \psi/\partial x)$, where the streamfunction $\psi$ reads:
\begin{equation}
\psi = \frac{1}{2}\Omega_0 |\mathbf x|^2 - \frac{\Gamma}{4 \pi} \sum_n \ln(|\mathbf x - \mathbf x_n|^2)
\label{streamfun}
\end{equation}
with $\mathbf x=(x,y)$. The corresponding streamlines for $N=$ 2, 3, 4 and 7 are shown in Fig.\ \ref{LdC_N2347}. In addition to the circular streamlines around vortices, the flow is   characterized by $N$ anticyclonic cells rotating around elliptic equilibrium positions of fluid points, that will be denoted $\mathbf X_{eq}^0$ throughout the paper (the superscript 0 indicating that they correspond to particles with $St = 0$).  
These anticyclonic cells will play a key role in the trapping of inertial particles: the Coriolis force acting on these particles will be directed towards the interior of the cell.
Note also that, as soon as $N >2$, another anticyclonic recirculation cell exists around the center-of-vorticity $(0,0)$, which is also an elliptic equilibrium position of fluid points. 
The equilibrium positions of fluid points  are fixed points of the system $ \dot {\mathbf X}=\mathbf u(\mathbf X) $. Their stability can be studied by means of the  eigenvalues $\mu$ of the gradient of $\mathbf u$, which satisfy
\begin{equation}
\mu^2 = (\frac{\partial u}{\partial x})^2 + (\frac{\partial u}{\partial y})^2 - 2  \frac{\partial u}{\partial y} \equiv I(\mathbf X).
\label{charpoly}
\end{equation}
Here we chose $1/\Omega_0$ as time unit,  and made use of the fact that the flow is incompressible and irrotational in the laboratory frame, that is $ \frac{\partial u}{\partial x} +  \frac{\partial v}{\partial y} = 0$ and $ \frac{\partial v}{\partial x} -  \frac{\partial u}{\partial y} = -2$. 
Using both expressions we could re-write the gradient of $\mathbf u$ in terms of derivatives of the $x$-velocity only, and obtain its characteristic polynomial (\ref{charpoly}).  
Elliptic equilibrium points $\mathbf X_{eq}^0$    are characterized by pure imaginary eigenvalues, so that
$I_0 \equiv I(\mathbf X_{eq}^0) < 0$.

In particular, the determinant of $\nabla \mathbf u$ at $\mathbf X_{eq}^0$ is equal to $|I_0|$ and is non-zero. The implicit functions theorem mentioned above implies the existence of  the branch $\mathbf X_{eq}(St)$, for inertial particles transported in this flow in the limit of small Stokes number $St$.
For a given Stokes number, the stability of the branch $\mathbf X_{eq}(St)$ emerging from 
$\mathbf X_{eq}^0$
can be investigated by linearising the particle motion equation (\ref{eqmvt}), in the vicinity of $(\mathbf X_{eq}(St),\dot{\mathbf X}=\mathbf 0)$, with $\mathbf F = \mathbf X - 2 \hat{\mathbf z} \times \dot{\mathbf X}$ corresponding to the non-dimensional centrifugal and Coriolis forces ($\hat{\mathbf z}$
is the  direct  unit vector perpendicular to the $(x,y)$ plane). The system is then recast as a dynamical system with 4 degrees-of-freedom $(\mathbf X,\dot{\mathbf X})$. 
Exploiting the fact that $St \ll 1$, one obtains that the eigenvalues of the dynamical system (\ref{eqmvt}) at the equilibrium point have the following real parts:
 \begin{eqnarray}
 \label{l1}
 Re(\lambda_1) &=& Re(\lambda_3) = - (1+I_0)\, St ,\\
 Re(\lambda_2) &=& Re(\lambda_4) = -\frac{1}{St} + (1+I_0) \, St 
 \label{l2}
 \end{eqnarray}
plus terms of order $O(St^2)$. Hence, if $1+I_0 \not = 0$, and even if the fluid stagnation point is degenerate (non-hyperbolic), the neighbouring  equilibrium point of inertial particles is not, since all real parts are non-zero in the limit of small Stokes numbers. It is hyperbolic and its stability can be derived by examining the sign of these real parts.
In the limit where $St \ll 1$, the real part of $\lambda_2$ and $\lambda_4$ is always strictly negative. If $1+I_0 > 0$, the real part of $\lambda_1$ and $\lambda_3$ is also strictly negative
and the equilibrium point $\mathbf X_{eq}(St)$ is asymptotically stable and attracts particles released in its basin of attraction. For the four cases of figure \ref{LdC_N2347}, the position of the fluid elliptic stagnation points in the anticyclonic cells could  easily be found (exactly for $N$=2 and 4), and one gets: $1+I_0= 1/4$ ($N=2$), 
$1+I_0 \simeq 0.121 $ ($N=3$), $1+I_0 = 11/4-\sqrt{7}$ ($N=4$), and $1+I_0 \simeq 0.335 $ ($N=7$). These values being positive, we conclude that inertial particles must be trapped in the vicinity of these points, provided their Stokes number is sufficiently small. Since the elliptic points in the anticyclonic cells are all located at $|\mathbf X| > 1$, that is outside the ring formed by the vortices, the  attracting  points $\mathbf X_{eq}(St)$ will be called {\it satellite}  attracting  points in the following.
 
In addition to these satellite   points, another equilibrium point exists for particles in these flows. Indeed, in the vicinity of the fluid stagnation point located at the center-of-vorticity $(0,0)$, one can check that the non-dimensional streamfunction reads $\psi \simeq \frac{1}{2}  r^2 + O(r^N)$, leading to $I_0 = -1$ when $N >2$. Eq.\ (\ref{l1})  therefore implies that 
$ Re(\lambda_1) = Re(\lambda_3) =   O(St^2)$, so that the linear analysis does not allow to conclude about the stability of  the equilibrium branch  near the center. Particles rotate around the center, while they can  drift towards the interior or the exterior of the cell under the effect of the non-linear terms $O(r^N)$ appearing in the streamfunction. To study the effect of these terms, we consider an inertial particle  injected very near $(0,0)$, i.e. $|\mathbf X(0)| = \varepsilon \ll 1$ and re-scale the positions  by $\varepsilon$, setting $x_* = x/\varepsilon$, $y_* = y/\varepsilon$. This leads to
\begin{equation}
\psi^*   = \frac{\psi}{\varepsilon^2} =  \psi_{0}^*+ \varepsilon^{N-2}  \psi_{N-2}^* + O(\varepsilon^{2N-2})
\end{equation}
where $ \psi_{0}^*=(x_*^2 + y_*^2)/2 $ and $\psi_{N-2}^*(x_*,y_*)$ is a polynomial of degree $N$: $\psi_{N-2}^*=x_*^3 - 3 x_* y_*^2$ (if $N=3$), $\psi_{N-2}^*=2/3\,{x_*}^{4}+2/3\,{y_*}^{4}-4\,{x_*}^{2}{y_*}^{2}$(if $N=4$), etc.
\begin{figure*}
\includegraphics[width=0.75\textwidth]{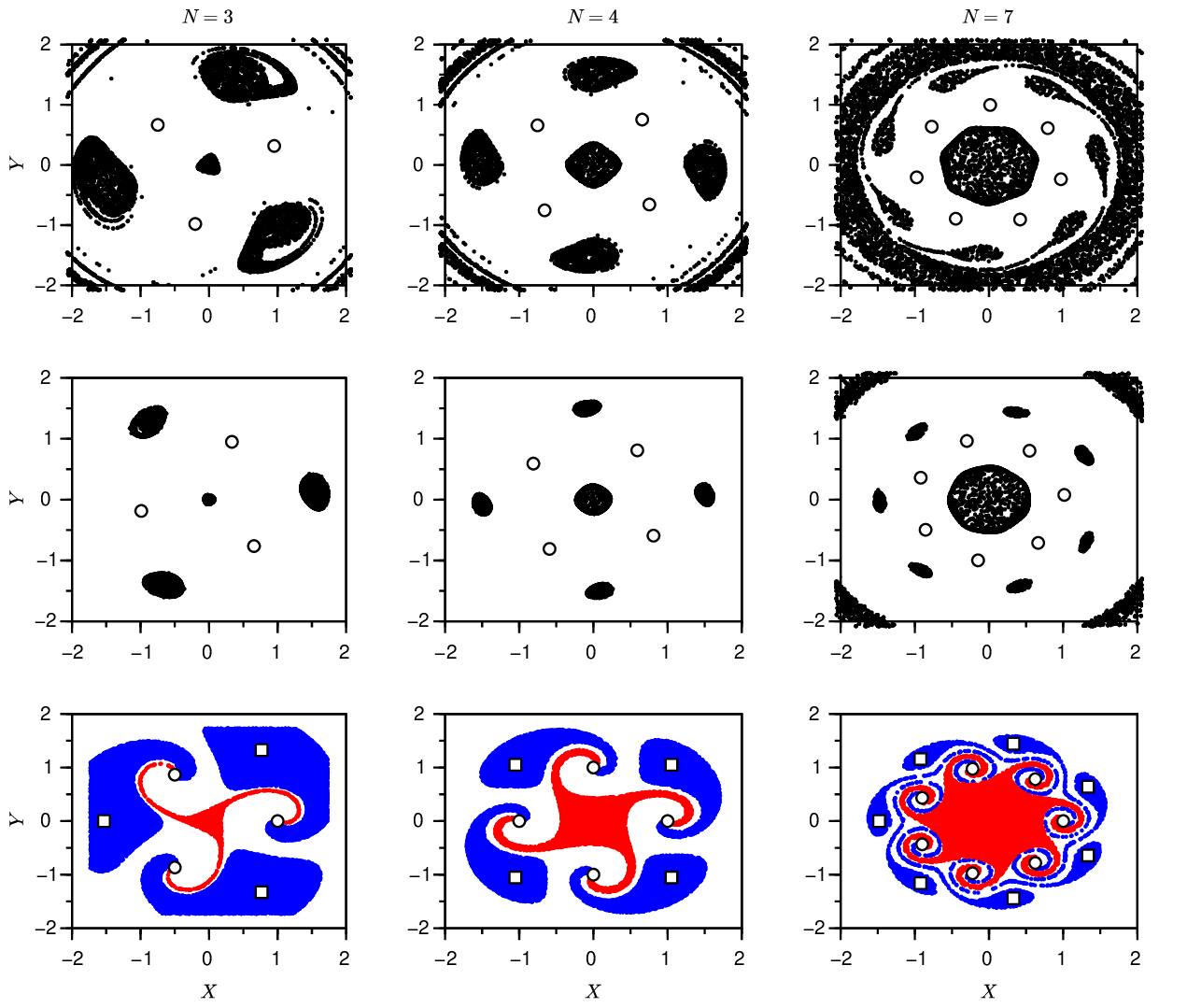}
\caption{Clouds of particles in the laboratory frame, advected by 3 vortices (first column), 4 vortices (middle column) and 7 vortices (third column), after two turns (upper row) and 12 turns (middle row) of the crystal.  White circles are vortex centers, and black dots are particles. 
The Stokes number of particles is $St=0.062$ for $N=3$ and 4 vortices, and $St=0.021$ for the case $N=7$. The last row shows the initial positions of the particles trapped by satellite points (blue dots) and by the center point (red dots). White squares indicate the position of the degenerate fluid   stagnation points $\mathbf X_{eq}^0$ in anticyclonic cells.}
\label{NuagesN347}       
\end{figure*}
Then we  calculate the change of the streamfunction
$\Delta \psi^* = \psi^*(\mathbf X(T))-\psi^*(\mathbf X(0))$ during one period $T= 2 \pi$ of the rotation of the crystal. Negative values of this quantity will indicate that the particle drifts towards the ring center (as $\psi^*$ increases with the distance to $(0,0)$, when $\varepsilon$ is sufficiently small).
To achieve this we consider a fixed value of $\varepsilon$ and assume that $\varepsilon^{2N-2} \ll St \ll \varepsilon$. On perturbing Eq.\ (\ref{eqmvt}) we get, for the re-scaled variables:
\begin{equation}
\dot {\mathbf X}_* \simeq \mathbf u_*  + St \, (-(\nabla \mathbf u_*)  \mathbf u_* + \mathbf X_* - 2 \hat{\mathbf z} \times  \mathbf u_*) + O(St^2)
\end{equation} 
where $\mathbf u_*  = \nabla \psi^* \times \hat{\mathbf z}$. We then obtain
$$
\Delta \psi^* = St \, \int_0^T \nabla \psi_*(\mathbf X(t))  \cdot \dot {\mathbf X}_* dt  
$$
\begin{equation}
= St \,\int_0^T [  \nabla \psi^*  \cdot   \mathbf X_* - 2   | \nabla \psi^* |^2  -\nabla \psi^*  \cdot (\nabla \mathbf u_*)  \mathbf u_*   ]dt .
\label{DeltaPsi}
\end{equation}
The first term in the integral (\ref{DeltaPsi}) is the contribution of the centrifugal force,
 the second one
is the contribution of the Coriolis force, and the last one is the contribution of the drag due to the  flow induced by vortices. 
As expected, the second term is always negative because the Coriolis force acts towards the right-hand-side of $\dot {\mathbf X}$ here, whereas the streamfunction $\psi^*$ increases towards the left-hand-side of streamlines. 
These integrals are calculated by approximating the particle trajectory  by the fluid point trajectory within integrands: $\mathbf X_*(t) = {\mathbf X}_f(t) + O(St)$ with $\dot {\mathbf X}_f \simeq \mathbf u_{*0} + \varepsilon^{N-2} \mathbf u^*_{N-2}$.  The solution is approximated in the form  ${\mathbf X}_f = (\cos t, -\sin t) + \varepsilon {\mathbf X}_1(t) + \varepsilon^2 {\mathbf X}_2(t)  + etc$, where the first term   corresponds to the motion of a fluid point when vortices are at infinity, and ${\mathbf X}_i(t)$    are to be found by injecting this expansion into the motion equation of fluid points.   In the case $N=3$, we get
\begin{equation}
\Delta \psi^* = -  36  \,\pi  \,St \, \varepsilon^2 
\label{DeltaPsiN3}
\end{equation}
plus terms of order $St^2$. We conclude that particles released in the vicinity of $\mathbf X=\mathbf 0$, i.e. within the anticyclonic cell around the center, will slowly drift towards the center. The set of vortices maintain the particles trapped in a kind of "cage". We note also that the drift of particles towards the center is very slow: back to the original scales (i.e. when lengths are reduced by the radius of the vortex crystal $a$), Eq.\ (\ref{DeltaPsiN3}) reads $\Delta \psi = - 36  \, \pi \,St  \, |\mathbf X(0)|^4$. Therefore, the change of streamfunction at each period is very small since $|\mathbf X(0)| \ll 1$. 
 
These theoretical results suggest  that particles, in a system of $N$ vortices located on a ring, can be trapped by $N$ satellite  attracting  points located outside the ring, and by an additional  attracting  point located at the center-of-vorticity. To check these results we have performed numerical simulations (in the laboratory frame with coordinates $(X,Y)$) involving $N$ point vortices moving under their mutual influence and initially placed on a ring with unit radius. In addition,  10000 inertial particles  have been released within the flow domain, and their positions have been computed by solving Eq.\ (\ref{eqmvt}).  It should be noted that when $N=3$ the critical Stokes number, below which equilibrium points do not exist, can be obtained analytically from Eq.\ (\ref{equil}). After a few manipulations using polar coordinates we show that the non-dimensional radius $r = |\mathbf X_{eq}| $ of equilibrium positions satisfies either $r=0$ or
$r^6 - {6} r^4 /(1+St^2)  +   {9} r^2 /(1+St^2)  - 1 = 0$.
This is a cubic equation in $r^2$, the discriminant of which is zero when the Stokes number is equal to the critical value:
  $$
  St_c = 1/2\,\sqrt {2\,{\frac {\sqrt {48\,A-{2}^{4/3} \, {19}^{2/3}A+280}}{
\sqrt {A}}}-4-2\,A} 
  $$
  where $A =  ({24+  {2}^{4/3} \, {19}^{2/3}})^{1/2}$,
leading to $St_c \approx 0.206$ in the three-vortex system,   in agreement with Fig.\ 3 of Ref. \cite{Ravichandran2017} (up to a $2\pi$ factor due to a different definition of the Stokes number).
For $N=4$ or more, no analytical expression can be found, and one has to use a numerical algorithm to approach $St_c$, leading to $St_c \approx 0.15$ ($N=4$) and $St_c \approx 0.05$ ($N=7$)   (see also Fig.\ 3 of Ref. \cite{Ravichandran2017}). The first column of Fig.\ \ref{NuagesN347} shows the resulting clouds of particles, when $N=3$ and $St=0.3 St_c$. It confirms the existence of the three satellite  attracting  points, as well as trapping near the center-of-vorticity. 
The second and third columns of Fig.\ \ref{NuagesN347} correspond to $N=4$ and $N=7$ respectively, with Stokes number chosen arbitrarily, such that $St < St_c$. As expected, some particles are trapped by satellite points, or by the center point.
Trapping by this point is very slow and long-term simulations (not shown here) would show tiny spots of particles near the satellite points, with a large cloud of particles slowly decaying near the center. The last row of Fig.\ \ref{NuagesN347} shows the initial positions of trapped particles, which give a view of the basin of attraction of the satellite points (blue dots), together with the basin of the center point (red dots) for $N=$ 3, 4 and 7.  The size of the basin of the central point significantly increases with $N$, as the size of the anticyclonic cell grows with $N$ and  most trapped particles come from the interior of this cell.    We note that these basins are not fractal, as expected since the dynamics is non-chaotic, but they have a spiral structure near vortices. Therefore, blue and red basins might be highly intricated near vortices.
   
   These analyses show  that inertial particles  might eventually form a mass crystal moving as a rigid-body with the vorticity crystal.   In addition, another  attracting  point exists near the center of the vortex crystal, in high contrast with the 2-vortex system where this point does not exist \cite{Angilella2010,Ravichandran2014}. These observations could reveal interesting features of particle dynamics in either laboratory of geophysical flows, where vortex crystals have been shown to emerge. Further analyses about particles lighter than the fluid, or with a density of the order of the fluid density, should also reveal interesting particle dynamics.  Finally, the case where a vortex is present at the center-of-vorticity, which is known to increase the stability of the crystal, is also  of interest for particle transport and will be the next step of this work.


\begin{thebibliography}{18}
\expandafter\ifx\csname natexlab\endcsname\relax\def\natexlab#1{#1}\fi
\expandafter\ifx\csname bibnamefont\endcsname\relax
  \def\bibnamefont#1{#1}\fi
\expandafter\ifx\csname bibfnamefont\endcsname\relax
  \def\bibfnamefont#1{#1}\fi
\expandafter\ifx\csname citenamefont\endcsname\relax
  \def\citenamefont#1{#1}\fi
\expandafter\ifx\csname url\endcsname\relax
  \def\url#1{\texttt{#1}}\fi
\expandafter\ifx\csname urlprefix\endcsname\relax\def\urlprefix{URL }\fi
\providecommand{\bibinfo}[2]{#2}
\providecommand{\eprint}[2][]{\url{#2}}

\bibitem[{\citenamefont{Aref et~al.}(2002)\citenamefont{Aref, Newton, Stremler,
  Tokieda, and Vainchtein}}]{Aref2002review}
\bibinfo{author}{\bibfnamefont{H.}~\bibnamefont{Aref}},
  \bibinfo{author}{\bibfnamefont{P.~K.} \bibnamefont{Newton}},
  \bibinfo{author}{\bibfnamefont{M.~A.} \bibnamefont{Stremler}},
  \bibinfo{author}{\bibfnamefont{T.}~\bibnamefont{Tokieda}}, \bibnamefont{and}
  \bibinfo{author}{\bibfnamefont{D.~L.} \bibnamefont{Vainchtein}},
  \bibinfo{journal}{TAM Reports 1008}  (\bibinfo{year}{2002}).

\bibitem[{\citenamefont{Yarmchuk et~al.}(1979)\citenamefont{Yarmchuk, Gordon,
  and Packard}}]{Yarmchuk1979}
\bibinfo{author}{\bibfnamefont{E.}~\bibnamefont{Yarmchuk}},
  \bibinfo{author}{\bibfnamefont{M.}~\bibnamefont{Gordon}}, \bibnamefont{and}
  \bibinfo{author}{\bibfnamefont{R.}~\bibnamefont{Packard}},
  \bibinfo{journal}{Physical Review Letters} \textbf{\bibinfo{volume}{43}},
  \bibinfo{pages}{214} (\bibinfo{year}{1979}).

\bibitem[{\citenamefont{Fine et~al.}(1995)\citenamefont{Fine, Cass, Flynn, and
  Driscoll}}]{Fine1995}
\bibinfo{author}{\bibfnamefont{K.}~\bibnamefont{Fine}},
  \bibinfo{author}{\bibfnamefont{A.}~\bibnamefont{Cass}},
  \bibinfo{author}{\bibfnamefont{W.}~\bibnamefont{Flynn}}, \bibnamefont{and}
  \bibinfo{author}{\bibfnamefont{C.}~\bibnamefont{Driscoll}},
  \bibinfo{journal}{Physical Review Letters} \textbf{\bibinfo{volume}{75}},
  \bibinfo{pages}{3277} (\bibinfo{year}{1995}).

\bibitem[{\citenamefont{Durkin and Fajans}(2000{\natexlab{a}})}]{Durkin2000PoF}
\bibinfo{author}{\bibfnamefont{D.}~\bibnamefont{Durkin}} \bibnamefont{and}
  \bibinfo{author}{\bibfnamefont{J.}~\bibnamefont{Fajans}},
  \bibinfo{journal}{Physics of fluids} \textbf{\bibinfo{volume}{12}},
  \bibinfo{pages}{289} (\bibinfo{year}{2000}{\natexlab{a}}).

\bibitem[{\citenamefont{Durkin and Fajans}(2000{\natexlab{b}})}]{Durkin2000PRL}
\bibinfo{author}{\bibfnamefont{D.}~\bibnamefont{Durkin}} \bibnamefont{and}
  \bibinfo{author}{\bibfnamefont{J.}~\bibnamefont{Fajans}},
  \bibinfo{journal}{Physical Review Letters} \textbf{\bibinfo{volume}{85}},
  \bibinfo{pages}{4052} (\bibinfo{year}{2000}{\natexlab{b}}).

\bibitem[{\citenamefont{Grzybowski et~al.}(2000)\citenamefont{Grzybowski,
  Stone, and Whitesides}}]{Grzybowski2000}
\bibinfo{author}{\bibfnamefont{B.~A.} \bibnamefont{Grzybowski}},
  \bibinfo{author}{\bibfnamefont{H.~A.} \bibnamefont{Stone}}, \bibnamefont{and}
  \bibinfo{author}{\bibfnamefont{G.~M.} \bibnamefont{Whitesides}},
  \bibinfo{journal}{Nature} \textbf{\bibinfo{volume}{405}},
  \bibinfo{pages}{1033} (\bibinfo{year}{2000}).

\bibitem[{\citenamefont{Grzybowski and Whitesides}(2002)}]{Grzybowski2002}
\bibinfo{author}{\bibfnamefont{B.~A.} \bibnamefont{Grzybowski}}
  \bibnamefont{and} \bibinfo{author}{\bibfnamefont{G.~M.}
  \bibnamefont{Whitesides}}, \bibinfo{journal}{The Journal of Physical
  Chemistry B} \textbf{\bibinfo{volume}{106}}, \bibinfo{pages}{1188}
  (\bibinfo{year}{2002}).

\bibitem[{\citenamefont{Adriani et~al.}(2018)\citenamefont{Adriani, Mura,
  Orton, Hansen, Altieri, Moriconi, Rogers, Eichst{\"a}dt, Momary, Ingersoll
  et~al.}}]{Adriani2018}
\bibinfo{author}{\bibfnamefont{A.}~\bibnamefont{Adriani}},
  \bibinfo{author}{\bibfnamefont{A.}~\bibnamefont{Mura}},
  \bibinfo{author}{\bibfnamefont{G.}~\bibnamefont{Orton}},
  \bibinfo{author}{\bibfnamefont{C.}~\bibnamefont{Hansen}},
  \bibinfo{author}{\bibfnamefont{F.}~\bibnamefont{Altieri}},
  \bibinfo{author}{\bibfnamefont{M.}~\bibnamefont{Moriconi}},
  \bibinfo{author}{\bibfnamefont{J.}~\bibnamefont{Rogers}},
  \bibinfo{author}{\bibfnamefont{G.}~\bibnamefont{Eichst{\"a}dt}},
  \bibinfo{author}{\bibfnamefont{T.}~\bibnamefont{Momary}},
  \bibinfo{author}{\bibfnamefont{A.~P.} \bibnamefont{Ingersoll}},
  \bibnamefont{et~al.}, \bibinfo{journal}{Nature}
  \textbf{\bibinfo{volume}{555}}, \bibinfo{pages}{216} (\bibinfo{year}{2018}).

\bibitem[{\citenamefont{Schecter et~al.}(1999)\citenamefont{Schecter, Dubin,
  Fine, and Driscoll}}]{Schecter1999}
\bibinfo{author}{\bibfnamefont{D.~A.} \bibnamefont{Schecter}},
  \bibinfo{author}{\bibfnamefont{D.~H.} \bibnamefont{Dubin}},
  \bibinfo{author}{\bibfnamefont{K.}~\bibnamefont{Fine}}, \bibnamefont{and}
  \bibinfo{author}{\bibfnamefont{C.}~\bibnamefont{Driscoll}},
  \bibinfo{journal}{Physics of Fluids} \textbf{\bibinfo{volume}{11}},
  \bibinfo{pages}{905} (\bibinfo{year}{1999}).

\bibitem[{\citenamefont{Jin and Dubin}(2000)}]{Jin2000}
\bibinfo{author}{\bibfnamefont{D.~Z.} \bibnamefont{Jin}} \bibnamefont{and}
  \bibinfo{author}{\bibfnamefont{D.}~\bibnamefont{Dubin}},
  \bibinfo{journal}{Physics of plasmas} \textbf{\bibinfo{volume}{7}},
  \bibinfo{pages}{1719} (\bibinfo{year}{2000}).

\bibitem[{\citenamefont{Siegelman et~al.}(2022)\citenamefont{Siegelman, Young,
  and Ingersoll}}]{Siegelman2022}
\bibinfo{author}{\bibfnamefont{L.}~\bibnamefont{Siegelman}},
  \bibinfo{author}{\bibfnamefont{W.~R.} \bibnamefont{Young}}, \bibnamefont{and}
  \bibinfo{author}{\bibfnamefont{A.~P.} \bibnamefont{Ingersoll}},
  \bibinfo{journal}{Proceedings of the National Academy of Sciences}
  \textbf{\bibinfo{volume}{119}}, \bibinfo{pages}{e2120486119}
  (\bibinfo{year}{2022}).

\bibitem[{\citenamefont{Angilella}(2010)}]{Angilella2010}
\bibinfo{author}{\bibfnamefont{J.~R.} \bibnamefont{Angilella}},
  \bibinfo{journal}{Physica D} \textbf{\bibinfo{volume}{239}},
  \bibinfo{pages}{1789} (\bibinfo{year}{2010}).

\bibitem[{\citenamefont{Ravichandran et~al.}(2014)\citenamefont{Ravichandran,
  Perlekar, and Govindarajan}}]{Ravichandran2014}
\bibinfo{author}{\bibfnamefont{S.}~\bibnamefont{Ravichandran}},
  \bibinfo{author}{\bibfnamefont{P.}~\bibnamefont{Perlekar}}, \bibnamefont{and}
  \bibinfo{author}{\bibfnamefont{R.}~\bibnamefont{Govindarajan}},
  \bibinfo{journal}{Physics of Fluids} \textbf{\bibinfo{volume}{26}}
  (\bibinfo{year}{2014}).

\bibitem[{\citenamefont{Ravichandran et~al.}(2017)\citenamefont{Ravichandran,
  Deepu, and Govindarajan}}]{Ravichandran2017}
\bibinfo{author}{\bibfnamefont{S.}~\bibnamefont{Ravichandran}},
  \bibinfo{author}{\bibfnamefont{P.}~\bibnamefont{Deepu}}, \bibnamefont{and}
  \bibinfo{author}{\bibfnamefont{R.}~\bibnamefont{Govindarajan}},
  \bibinfo{journal}{S{\=a}dhan{\=a}} \textbf{\bibinfo{volume}{42}},
  \bibinfo{pages}{597} (\bibinfo{year}{2017}).

\bibitem[{\citenamefont{Babiano et~al.}(2000)\citenamefont{Babiano, Cartwright,
  Piro, and Provenzale}}]{Babiano2000}
\bibinfo{author}{\bibfnamefont{A.}~\bibnamefont{Babiano}},
  \bibinfo{author}{\bibfnamefont{J.}~\bibnamefont{Cartwright}},
  \bibinfo{author}{\bibfnamefont{O.}~\bibnamefont{Piro}}, \bibnamefont{and}
  \bibinfo{author}{\bibfnamefont{A.}~\bibnamefont{Provenzale}},
  \bibinfo{journal}{Phys. Rev. Let.} \textbf{\bibinfo{volume}{84}},
  \bibinfo{pages}{5764} (\bibinfo{year}{2000}).

\bibitem[{\citenamefont{Haller and Sapsis}(2008)}]{Haller2008}
\bibinfo{author}{\bibfnamefont{G.}~\bibnamefont{Haller}} \bibnamefont{and}
  \bibinfo{author}{\bibfnamefont{T.}~\bibnamefont{Sapsis}},
  \bibinfo{journal}{Physica D} \textbf{\bibinfo{volume}{237}},
  \bibinfo{pages}{573} (\bibinfo{year}{2008}).

\bibitem[{\citenamefont{Cartwright et~al.}(2010)\citenamefont{Cartwright,
  Feudel, Karolyi, De~Moura, Piro, and Tel}}]{Cartwright2010}
\bibinfo{author}{\bibfnamefont{J.}~\bibnamefont{Cartwright}},
  \bibinfo{author}{\bibfnamefont{U.}~\bibnamefont{Feudel}},
  \bibinfo{author}{\bibfnamefont{G.}~\bibnamefont{Karolyi}},
  \bibinfo{author}{\bibfnamefont{A.}~\bibnamefont{De~Moura}},
  \bibinfo{author}{\bibfnamefont{O.}~\bibnamefont{Piro}}, \bibnamefont{and}
  \bibinfo{author}{\bibfnamefont{T.}~\bibnamefont{Tel}}, pp.
  \bibinfo{pages}{51--87} (\bibinfo{year}{2010}), \bibinfo{note}{non-linear
  Dynamics and Chaos: Advances and Perspectives. Thiel, M. Ed. Springer-Verlag
  Berlin Heidelberg.}

\bibitem[{\citenamefont{Mertz}(1978)}]{Mertz1978}
\bibinfo{author}{\bibfnamefont{G.~J.} \bibnamefont{Mertz}},
  \bibinfo{journal}{The Physics of Fluids} \textbf{\bibinfo{volume}{21}},
  \bibinfo{pages}{1092} (\bibinfo{year}{1978}).

\end{thebibliography}
\end{document}